\documentclass[letterpaper,10pt]{article} 

\usepackage{tabularx}
\newcolumntype{Y}{>{\centering\arraybackslash}X}
\usepackage{pks} 

\usepackage{caption}
\usepackage{multicol}
\newcommand\authormark[1]{\textsuperscript{#1}}
\usepackage{amsmath,amssymb}
\usepackage[colorlinks=true,bookmarks=false,citecolor=blue,urlcolor=blue]{hyperref} 

\begin{document}

\title{Accurate and Fast VR Eye-Tracking using Deflectometric Information}

\vspace{-10px}

\author{Jiazhang Wang\authormark{1,*}, Tianfu Wang\authormark{2,*}, Bingjie Xu\authormark{3}, \\ Oliver Cossairt\authormark{1,3} and Florian Willomitzer\authormark{4,\dag}}

\address{\authormark{1} Department of Electrical and Computer Engineering, Northwestern University, Evanston, IL, 60208\\
\authormark{2}Department of Computer Science, ETH Zürich, Zürich, Switzerland, 8092\\
\authormark{3} Department of Computer Science, Northwestern University, Evanston, IL, 60208\\
\authormark{4}Wyant College of Optical Sciences, University of Arizona, Tuscon, AZ, 85721\\
\authormark{*}These two authors contributed equally}
\vspace{-5px}
\email{\authormark{\dag}fwillomitzer@arizona.edu } 

\vspace{-18px}

\begin{abstract}
We present two methods for fast and precise eye-tracking in VR headsets. Both methods exploit deflectometric information, i.e., the specular reflection of an extended screen over the eye surface. 
\end{abstract}

\vspace{-3px}

\section{Introduction}
A fast and accurate solution to eye-tracking is vital for many functions in Virtual Reality(VR) headsets, including foveated rendering, virtual avatar interaction, or increasing the viewing comfort. Most state-of-the-art eye-tracking methods either exploit 2D features detected from 2D eye images (``image-based methods''), or use sparse reflections of a few point light sources at the eye surface (``reflection-based methods''). A prominent example of the latter is ``glint tracking'' which samples the eye surface  at $\sim$ 10-15 point source reflections. For those state-of-the-art methods, the density of measured surface points (2D features for image-based methods, point source reflections for glint tracking) is relatively low, and the acquired 3D information about the eye surface is limited \cite{wang2021vr, wang2023optimization}.

In \cite{wang2021vr} we introduced a novel concept for eye tracking in VR headsets that utilizes Deflectometry\cite{willomitzer2020hand} to estimate the gaze direction: The specular reflection of an extended screen displaying a known pattern is observed over the eye surface. By observing the deformation of the pattern in the camera image, dense 3D information about the eye surface (such as surface shape and normal) can be extracted, which is then used to estimate the gaze direction. The acquired data density is significantly higher than the density of the sparse methods discussed above. Factors of 1000X and higher are easily achievable. We refer to \cite{wang2021vr} for more information.

In this contribution, we present two novel methods based on our original idea described above. One method utilizes single-shot stereo Deflectometry for fast and precise eye surface measurement; the other method uses an optimization-based inverse rendering approach to estimate the gaze direction from the captured deflectometric information \cite{wang2023optimization}. For both methods, we show quantitative gaze evaluation results from real world experiments.
\vspace{-2px}
\begin{figure}[b]
  \centering
  \includegraphics[width=0.8\linewidth]{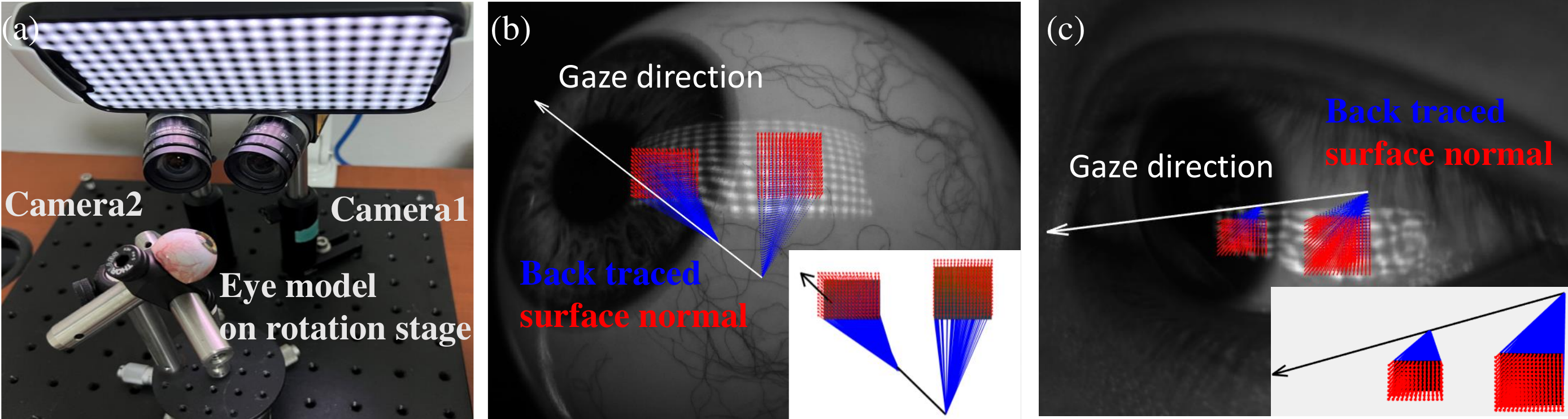}
\captionsetup{font=footnotesize}
\caption{\textbf{Eye tracking using single-shot stereo Deflectometry.}(a) Prototype setup. (b,c) Captured camera images of realistic eye model and real human eye, overlayed with the calculated surface normals and estimated gaze direction.}
\label{F1}
\end{figure}


\section{Methods and Results}

\textbf{Single-shot stereo deflectometry approach:} This method uses a crossed fringe pattern on the screen to measure the surface normal map of the eye surface. To establish the required correspondence between screen and camera, the phase information for both fringe directions is evaluated \textit{in single-shot} via a 2D continuous wavelet transform approach~\cite{liang2020using}. The second camera is used to solve the Deflectometry normal-depth ambiguity problem \cite{knauer2004phase}. This allows for a unique reconstruction of the shape \textit{and} normal map of the eye surface in single-shot. To estimate the gaze direction, the calculated surface normals are traced back towards the eye center. Due to the spherical shape of the eye's cornea and sclera,  but their  vastly different radii, the back-traced normals intersect at two different points inside the eye (cornea and sclera center). Connecting these two points delivers the optical axis of the eye and the gaze direction. We note that this approach also works if cornea and sclera are not perfectly spherical but rotational symmetric. In this case, all back-traced normals intersect along a line that coincides with the optical axis of the eye.

We evaluate this method by performing a quantitative real-world experiment on a realistic eye model mounted on a rotation stage (see setup in Fig.~\ref{F1}a). We rotate the eye model to different rotation positions~$a$ ($-3^\circ$, $0^\circ$,   $3^\circ$, and $6^\circ$) and evaluate the gaze angle $\theta_{a}$ as described above. Since the absolute gaze angle is unknown, we calculate the \textit{relative gaze angle} between two rotation positions and compare the result with the angle we rotated the eye model (``ground truth''). We acquired  20 measurements for each rotation position (80 measurements in total), while the eye model was always moved/rotated before a measurement was taken. Eventually, we evaluated the mean relative error $\epsilon_{0^\circ}$ at the rotation position $0^\circ$ w.r.t. all other rotation positions $a$: $\epsilon_{0^\circ}= ||\bar{\theta_{a}} - \bar{\theta_{0^\circ}}| - |a - 0^\circ||$. The results are shown in Tab.\ref{t1} left. It can be seen that $\epsilon_{0^\circ}$ is below $0.35^\circ$ for all measurements. To show the feasibility of our method under real conditions, we also performed a first qualitative single-shot measurement on a real living human eye. The result is shown in Fig.\ref{F1}.(c). It can be seen that the measurement on the human eye  shows the same properties as the measurement on the model eye: The evaluated surface normals intersect at two points and the evaluated gaze vector is qualitatively correct. A procedure for the quantitative evaluation of the absolute gaze direction is still current work in progress.

\begin{table}[h!]
\begin{center}
\begin{tabular}{|c|p{1.25cm}|p{1.25cm}|p{1.25cm}|p{1.0cm}|p{1.0cm}|p{1.0cm}|p{1.0cm}|}
\hline
\multicolumn{1}{|c|}{Method} & \multicolumn{3}{|c|}{Single-shot stereo Deflectometry}  & \multicolumn{4}{|c|}{Optimization-based inverse rendering} \\ \hline
Rotation position $a$  & -3$^\circ$   & 3$^\circ$    & 6$^\circ$    & -4$^\circ$   & -2$^\circ$   & 2$^\circ$    & 4$^\circ$   \\ \hline
Mean relative error $\epsilon_{0^\circ}$ & 0.34$^\circ$ & 0.24$^\circ$ & 0.17$^\circ$ & 0.45$^\circ$ & 0.27$^\circ$ & 0.28$^\circ$ & 0.33$^\circ$  \\ \hline
\end{tabular}
\vspace{-5px}
\captionsetup{font=footnotesize}
\caption{Evaluation of estimated gaze direction for single-shot stereo Deflectometry method and optimization-based method}
\label{t1}
\end{center}
\end{table}
\vspace{-18px}

\begin{figure}[htbp]
  \centering
  \includegraphics[width=0.8\linewidth]{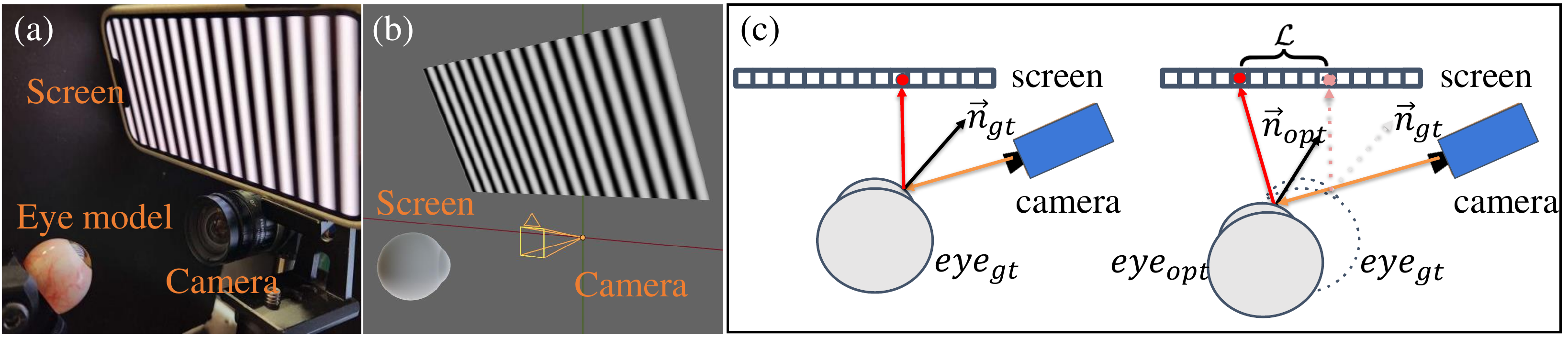}
\vspace{-10px}
\captionsetup{font=footnotesize}
\caption{\textbf{Optimization-Based Eye Tracking using Deflectometric Information}(a) Prototype setup. (b) ``Digital copy'' of the prototype setup used for differentiable rendering pipeline. (c) Obtaining the eye's rotation, translation, and shape over screen-camera correspondences via differential rendering (see also \cite{wang2023optimization}).}
\label{F2}
\end{figure}

\noindent
\textbf{Optimization-based inverse rendering approach:} This method uses the known geometry of our calibrated\cite{wang2022easy} deflectometric setup (see Fig.\ref{F2}(a)) to develop a PyTorch3D-based differentiable rendering pipeline that simulates a virtual computer-generated (CG) eye model under screen illumination (Fig.\ref{F2}.(b)). Eventually, the images and screen-camera correspondence information of the \textit{real} eye measurement is used to optimize the CG eye’s rotation, translation, and shape parameters with our renderer via gradient descent. This is done until the simulated setup with the CG eye produces images and correspondences that closely match the real measurements, and the gaze direction of the CG eye is eventually used as an estimate of the real eye's gaze direction. The method does not require a second camera. Moreover, it does not require a specific screen pattern and can even work with ordinary video frames of the main VR screen itself - also in single-shot. We refer to \cite{wang2023optimization} for more information. We evaluated our optimization-based method with the same experiment described above at different rotation positions ($-4^\circ$,$-2^\circ$, $0^\circ$,   $2^\circ$, and $4^\circ$), using phase shifted sinusoids as screen pattern. The evaluated mean relative error $\epsilon_{0^\circ}$ at the rotation position $0^\circ$ w.r.t. all other rotation positions $a$ is given in Tab.\ref{t1} right. All evaluated errors $\epsilon_{0^\circ}$ are below $0.5^\circ$
\vspace{-5px}
\vspace{-20px}

\bibliographystyle{unsrt}
\bibliography{sample}

\end{document}